\documentclass[useAMS,usenatbib]{mn2e}
\pdfoutput=1
\usepackage{times}
\usepackage{graphicx}
\usepackage{subfigure}
\usepackage{amsmath}
\usepackage{ifpdf}

\ifpdf
  \usepackage[pdftex, colorlinks = true, linkcolor = black,%
            plainpages = false, citecolor= black, 
            pagecolor = black, pdfpagelabels]{hyperref}
\else

\fi

\title[Efficient multimodal inference]{Efficient Bayesian inference for multimodal problems in cosmology}
\author[J.R. Shaw, M. Bridges and M.P. Hobson]
  {J.R. Shaw,$^1$\thanks{E-mail: jrs65@ast.cam.ac.uk}
   M. Bridges,$^2$ M.P. Hobson$^2$\\
  $^1$Institute of Astronomy, Madingley Road, Cambridge, CB3 0HA, UK\\
  $^2$Astrophysics Group,
      Cavendish Laboratory, Madingley Road,
      Cambridge CB3 0HE, UK\\
}

\date{Accepted 2007 April 16. Received 2007 April 5; in original form \today}
\pagerange{\pageref{firstpage}--\pageref{lastpage}}
\pubyear{2006}

\begin{document}
\label{firstpage}
\maketitle

\begin{abstract}
  Bayesian model selection provides the cosmologist with an exacting
  tool to distinguish between competing models based purely on the
  data, via the Bayesian evidence. Previous methods to calculate this
  quantity either lacked general applicability or were computationally
  demanding. However, nested sampling (Skilling 2004), which was
  recently applied successfully to cosmology by Muhkerjee et al. 2006,
  overcomes both of these impediments. Their implementation restricts
  the parameter space sampled, and thus improves the efficiency, using
  a shrinking ellipsoidal bound in the $n$-dimensional parameter space
  encompassing parameter samples above a decreasing likelihood value.
  However, if the likelihood function contains any multi-modality,
  separated over a significant portion of the parameter space then the
  ellipse is prevented from constraining the sampling region by less
  than the distance between the likelihood peaks. In this paper we
  introduce a method of clustered nested sampling whereby ellipsoidal
  clusters can form on any peaks identified --thus improving the
  efficiency by a factor which is equal to the ratio of the volumes
  enclosed by the set of small clustered ellipsoids and the large
  single ellipse that would necessarily be required without
  clustering. In addition we have implemented a method for determining
  the expectation \emph{and} variance of the final evidence value
  without the need to use sampling error from repetitions of the
  algorithm ; this further reduces the computational load by at least
  an order of magnitude. We have applied our algorithm to a pair of
  toy models and one cosmological example where we demonstrate that
  the number of likelihood evaluations required is $\sim$ 4$\%$ of
  that necessary for using previous algorithms.  We have produced a
  {\sc Fortran} library containing our routines which can be called
  from any sampling code, in addition for convenience we have
  incorporated it into the popular {\sc CosmoMC} code as {\sc
    CosmoClust}. Both are available for download at {\tt
    www.mrao.cam.ac.uk/software/cosmoclust}.
\end{abstract}


\section{Introduction} 

% Bayesian inference has become an instrumental tool in analysing
% cosmological datasets.  Until recently it has been solely applied to
% the generation of the posterior probability distribution and thus to
% place constraints on model parameters. The more fundamental question
% of model selection has been sidelined due to its heavy computational
% requirements. Ideally of course one would like to know whether a
% model was worth considering before calculating parameter
% constraints.

Bayesian inference has become an invaluable tool in analysing
cosmological datasets to place constraints on model
parameters. However the more fundamental question of model selection,
naturally incorporated in the Bayesian framework by the evidence, has
been under-utilised in cosmology.  Impeded until recently by the high
computational cost of methods such as thermodynamic integration,
progress has now been made with the advent of nested sampling
\citep{Skilling}, targeted solely at efficient calculation of the
evidence but also capable of providing posterior inferences. This
method has recently been applied to cosmological problems by
\citet{Mukherjee} allowing model selection on feasible timescales,
typical of Markov Chain Monte Carlo (MCMC) parameter estimation. Their
algorithm uses an elliptical bound to restrict the prior around the
maximum likelihood peak in the posterior and thus improves the
acceptance ratio and efficiency. While this method can be used for the
majority of cosmological applications, where posteriors are generally
uni-modal, there are examples where this is not so. In this paper we
describe a \emph{clustering} nested sampler which is capable of
detecting and isolating multiple peaks in the posterior, fitting
separate ellipsoidal bounds around each and making considerable
savings in computational load when compared to a single ellipse. In
addition we have implemented an improved error calculation
\citep{Skilling} on the final evidence result which produces a mean
and standard error in one calculation, eliminating the need for
multiple runs.

\section{Bayesian Inference}

A Bayesian analysis provides a coherent approach to the estimation of
a set of model parameters $\mathbf{\Theta}$ and, crucially in
determining which model, $M$, best describes the data, $\mathbf{D}$.
Bayes' theorem states that
\begin{equation} P(\mathbf{\Theta}|\mathbf{D}, M) =
\frac{P(\mathbf{D}|\mathbf{\Theta},
M)P(\mathbf{\Theta}|M)}{P(\mathbf{D}|M)},
\end{equation}
where $P(\mathbf{\Theta}|\mathbf{D}, M)$ is the posterior probability
distribution, $P(\mathbf{D}|\mathbf{\Theta}, M) \equiv \mathcal{L}$ the
likelihood, $P(\mathbf{\Theta}|M) \equiv \pi$ the prior, $P(\mathbf{D}|M) \equiv
\mathcal{Z}$ the Bayesian evidence.

Typically, the posterior is generated via a Metropolis-Hastings
algorithm with MCMC sampling, where at equilibrium the chain contains
a set of samples from parameter space distributed with the posterior
probability distribution. In the Bayesian framework all of the
inference is contained in the final multi-dimensional posterior, which
can be marginalised over each parameter to obtain constraints. The
Bayesian evidence is represented by the overall normalisation of this
posterior. The evidence is large in models which have a high
likelihood over a large proportion of the prior parameter space so
that we can consider the evidence to be the average of the likelihood
divided by the prior. As such it forms the integral, over the
$n$-dimensional parameter space
\begin{equation}
\mathcal{Z} = \int{\mathcal{L}(\mathbf{\Theta})\pi(\mathbf{\Theta})}d^n\mathbf{\Theta}.
\label{equation:evidence}
\end{equation} 
Thus a model containing a large number of parameters for which only a
narrow region of the prior is likely will have a low evidence and vice
versa, providing a natural mechanism to limit the complexity of
cosmological models and elegantly incorporating Ockham's razor.

A standard scenario in Bayesian model selection would require the
computation of evidences for two models A and B. The difference of
log-evidences $\ln \mathcal{Z}_A - \ln \mathcal{Z}_B$, also called the
Bayes factor then quantifies how well A may fit the data when compared
with model B. \citet{Jeffreys} provides a scale on which we can make
qualitative conclusions based on this difference: $\Delta\mbox{ln} \mathcal{Z} <
1$ is not significant, $1 < \Delta\mbox{ln} \mathcal{Z} < 2.5$ significant, $2.5
< \Delta\mbox{ln} \mathcal{Z} < 5$ strong and $\Delta\mbox{ln} \mathcal{Z} > 5$ decisive.

%% Modification: Noted that it's not 10^7, but 10^6 per chain, with %% around 10 chains needed. 
The prior dependence of the evidence requires that the entire
parameter space is adequately sampled, the MCMC method described above
moves rapidly through parameter space toward areas of high likelihood
leaving the majority of the surrounding prior heavily under
sampled. This is sufficient to generate parameter constraints where
regions close to the peak of the posterior are most important, but it
radically over estimates the evidence due to under sampling of regions
of low likelihood. In order to sample more uniformly the method of
thermodynamic integration has previously been implemented (see e.g
\citealt{Hobson}; \citealt{Slosar}) and used successfully by a number
of authors (\citealt{Niarchou}, \citealt{Basset}, \citealt{Mukherjee},
\citealt{Trotta}, \citealt{Beltran}, \citealt{Bridges}) in computing
the evidence.  Thermodynamic integration uses MCMC to draw samples not
from the posterior directly but from $\mathcal{L}^{\lambda}\pi$ where
$\lambda$ is the inverse temperature and is raised from $\approx 0$ to
$1$. For low values of $\lambda$ peaks in the posterior are
sufficiently flattened to allow improved mobility of the chains over
the entire prior range. Typically it is possible to obtain accuracies
of within 0.5 units in log-evidence via this method however this does
require of order $10^6$ samples per chain (with around 10 chains
required to determine a sampling error) making it at least an order of
magnitude more costly than the sampling needed for parameter
estimation.

\section{Nested sampling}
\label{section:nested} 

\begin{figure}
 \begin{center}
    	\subfigure[]{
          \includegraphics[width=0.4\columnwidth]{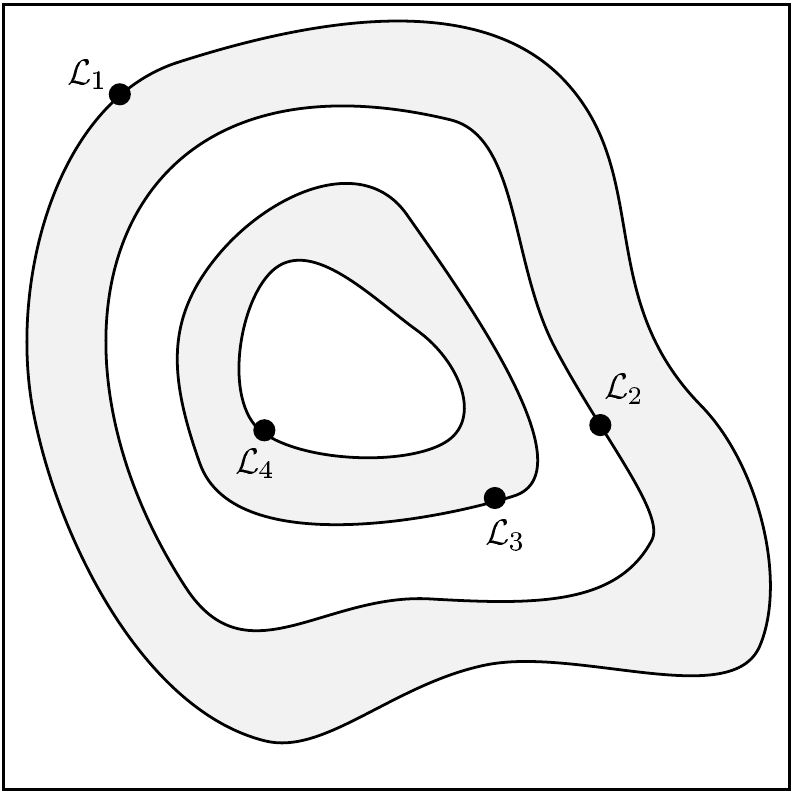}}
	  \hspace{0.3cm}
	\subfigure[]{
          \includegraphics[width=0.4\columnwidth]{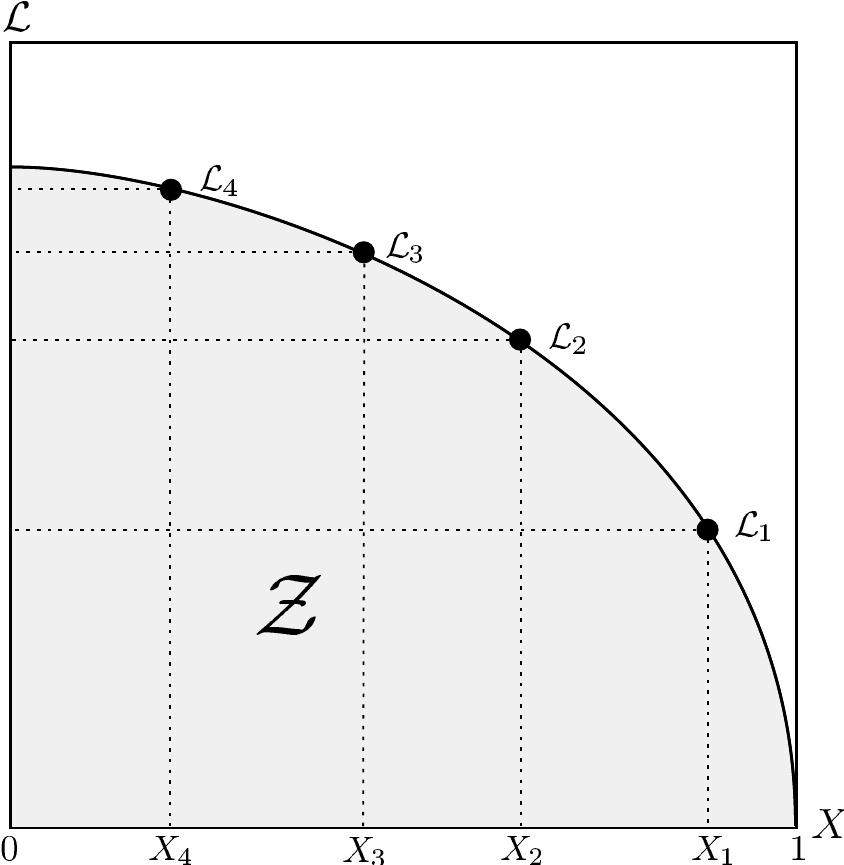}}
\caption{Cartoon illustrating the posterior of a two dimensional problem (a)
and the transformed $\mathcal{L}(X)$ function (b) where the prior masses
$X_i$ are
associated with each likelihood $\mathcal{L}_i$.}
\label{figure1}
\end{center}
\end{figure}

%% Changed to X \in [0,1]
Nested sampling is computationally more efficient as it transforms the
integral in Eqn. \ref{equation:evidence} to a single dimension by a
suitable re-parameterisation in terms of the prior \emph{mass} $X$. This mass can be divided
into elements $dX = \pi(\mathbf{\Theta})d^N \mathbf{\Theta}$ which can
be combined in any order to give say
%%% Changed integral to "L(\Theta) > \Lambda"
\begin{equation}
X(\lambda) = \int_{\mathcal{L\left(\mathbf{\Theta}\right) > \lambda}} \pi(\mathbf{\Theta}) d^N
\mathbf{\Theta},
\end{equation}
the prior mass covering all likelihoods above the iso-likelihood curve
$\mathcal{L} = \lambda$.  We also require the function
$\mathcal{L}(X)$ to be a singular decreasing function (which is
trivially satisfied for most posteriors) so that using sampled points
we can estimate the evidence via the integral:
\begin{equation}
\mathcal{Z}=\int_0^1{\mathcal{L}(X)}dX.
\label{equation:nested}
\end{equation}
An example of a posterior in two dimensions and its associated
function $\mathcal{L}(X)$ is shown in Fig.  \ref{figure1}.

%%% I've reordered the next few paragraphs such that they read through
%%% in a better order.

The transformed integral is dominated by a relatively small region of
the prior containing the majority of the posterior mass. For
efficient sampling the division of the prior elements should not be
linear but geometric:

%
%%% Correction.
%%% This doesn't make any sense, though I may be being stupid.
% \begin{equation}
%  X_0 = 1, X_1 = t_1 \frac{X_1}{X_{0}} \cdot \cdot \cdot, X_i = t_i \frac{X_i}{X_{i-1}},
%  \cdot\cdot\cdot, X_m = 0
% \label{equation:recursion}
% \end{equation}
%
%%% This does.
\begin{equation}
 X_0 = 1, X_1 = t_1 X_{0},\: \ldots\: X_i = t_i X_{i-1},\: \ldots\: X_m = 0
\label{equation:recursion}
\end{equation}
where we would desire a largely constant $t_i$; the procedure for
nested sampling ensures just that.

% Changed interval limits below [0,X_1]
The method is then to draw an initial set of $N$ \emph{active} samples
uniformly from the full prior $[0,X_0 = 1]$. The samples are ordered in
terms of likelihood, the smallest of which, $\mathcal{L}_0$ is removed
from the active set. The prior is then reduced according to
Eqn. \ref{equation:recursion} so that the region sampled becomes
$[0,X_1]$ where $X_1= t_1 X_{0}$, a point is then found to replenish
the active set subject to the criterion that its likelihood
$>\mathcal{L}_0$. This sampling/replenishment cycle is then repeated until
the entire prior has been traversed. The algorithm thus travels through nested
shells of likelihood as the prior mass is reduced.  At this point
the evidence is approximated by a numerical
integration routine such as the trapezoid rule:
\begin{equation}
\mathcal{Z} = \sum_{i=1}^{N} \mathcal{L}_i \frac{(X_{i-1}-X_{i+1})}{2}.
\end{equation}
The method described above ensures that the $t_i$ have the
distribution $P(t_i) = Nt_i^{N-1}$, allowing them to be statistically assigned. This
 does introduce an uncertainty, but it is possible to quantify this accurately and relate it to 
 the final $\mathcal{Z}$.
If we could assign each $X_i$ exactly then the only error would be due
to the discretisation of the integral which, given sufficient points
would be expected to be negligible. For this reason the dominant
source of error in the final $\mathcal{Z}$ arises from the
incorrect assignment of each prior mass. Fortuitously our knowledge of
the distribution $P(t_i)$ from which each $t_i$ is drawn allows us to
assess the errors in any quantities we produce.  Given the probability
of a vector $\mathbfss{t}$ as
\begin{equation}
P(\mathbfss{t}) = \prod_i P(t_i)
\end{equation}
we can write the expectation value of some quantity $F(\mathbfss{t})$ as
\begin{equation}
\langle F \rangle = \int F(\mathbfss{t})P(\mathbfss{t}) d^M\mathbfss{t}.
\end{equation}
Evaluating this integral is possible by Monte Carlo methods by
sampling a given number of vectors $\mathbfss{t}$ and finding the
average $F$. By this method we can essentially determine the variance of the
curve in $X-\mathcal{L}$ space, and thus in the evidence integral
$\int \mathcal{L}(X)dX$.

Previous numerical evidence analyses have determined an associated
error on the final evidence estimate by repeating identical
calculations to determine the standard error, and for reliable
results this method requires of order tens of estimates. As our
results will show the above method is capable of estimating the
expectation and variance of $\mathcal{Z}$ accurately when compared to
sampling statistics so as to eliminate the need for any repetition.

\subsection{Ellipsoidal Sampling}
Evaluating the likelihood of a point in cosmological parameter space
is computationally expensive. Nested sampling reduces this overhead as
it requires fewer likelihood evaluations. However further reductions
are possible by restricting the prior region at successive
sample/replacement cycles. Drawing blindly from the prior probability
distribution, will invariably result in an increasing number of
samples from unlikely regions of the prior. Ellipsoidal sampling
partially overcomes this by approximating the iso-likelihood of the
point to be replaced by an $n$-dimensional ellipsoid formed from the
covariance matrix of the current set of active points. In the limit as
the ellipsoid approaches that of the likelihood bound, the acceptance
rate tends to unity.

\begin{figure}
%\psfrag{labelx}{$en$}
%\psfrag{labely}{$\ln \mathcal{Z}$} 
 \begin{center}
          \includegraphics[width=0.7\columnwidth, angle = -90]{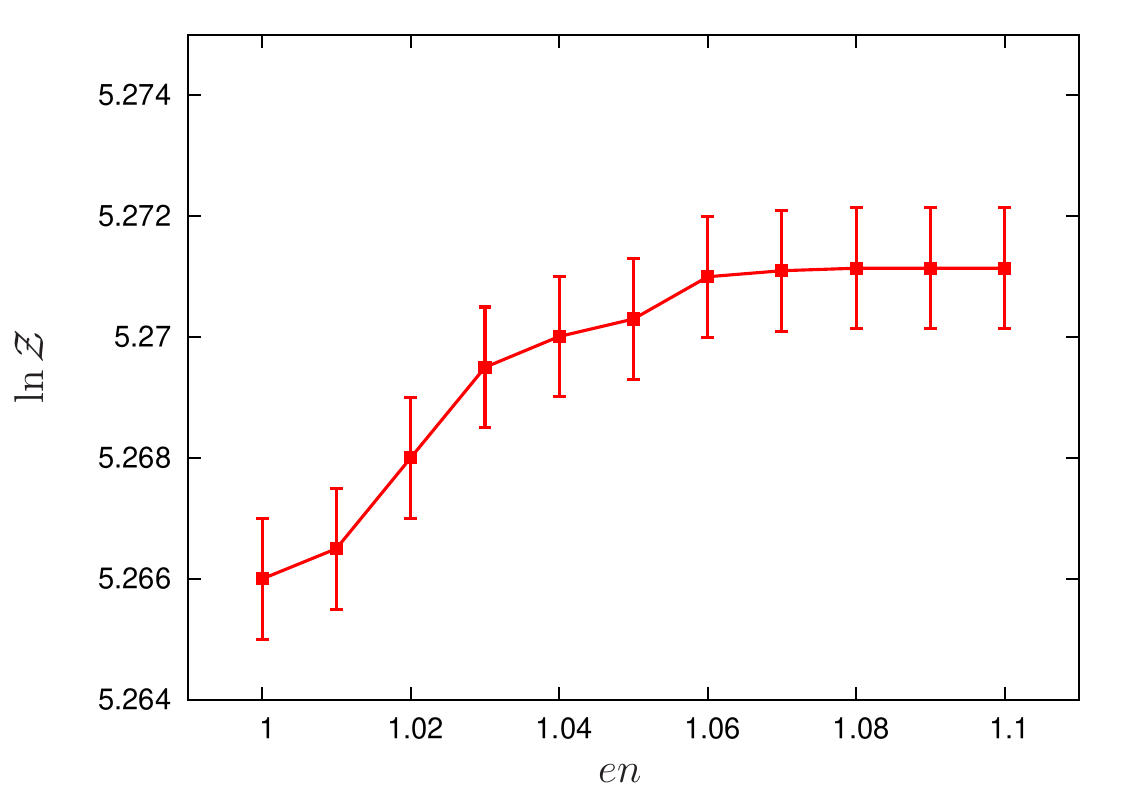}
\caption{$\ln \mathcal{Z}$ as determined for a range of enlargement factors for a multiple Gaussian posterior, the
analytic value in 5.271.}
\label{enlargement}
\end{center}
\end{figure}

%
%% Modification: the comment on Mukherjee's scaling stuff, I'm not
%% sure about "..on a very degenerate posterior..", the simplest
%% example I have is something rectangular.
Ellipsoidal sampling provides an elegant solution, but there are
some algorithmic points to note. Firstly, it is imperative that
the ellipse does not restrict the sampling region to lie within
$\mathcal{L}=\mathcal{L}_{\rm i}$ to avoid overestimating the
evidence. To prevent this, previous authors (such as \citealt{Mukherjee})
enlarged the ellipse by a fixed enlargement factor $ > 1$, so that in
most, but not all cases, the ellipse will encompass all active points
at any stage. An obvious example where this may break down would be a
rectangular posterior. We have chosen instead to define the ellipse to
guarantee that at any stage all active points are enclosed. If we have
determined the covariance matrix $\mathbfss{C}$ of samples at a given
stage, the diagonal matrix $\mathbfss{C}'$ is formed by transformation
$\mathbfss{C}'=\mathbfss{X}^{\mathbfss{T}}\mathbfss{C}\mathbfss{X}$
where $\mathbfss{X}$ is the matrix of eigenvectors of $\mathbfss{C}$,
where each eigenvalue is the squared length of the ellipse along each
principle axis. Thus the ellipsoidal bound is defined as
\begin{equation}
\mathbf{x}^{\mathbf{T}}\mathbf{C}^{-1}\mathbf{x} = k,
\end{equation}
where $\mathbf{x}$ are the parameter vectors along each axis and
where the constant scaling length
\begin{equation}
k = \max [\mathbf{x}_{\rm i}^{\mathbfss{T}} \mathbfss{C}^{-1} \mathbf{x}_{\rm i}]
\end{equation}
with $\mathbf{x}_{\rm i}$ denoting the vectors of the active point
set. This ensures that all active points are enclosed within the
ellipse although there does, of course, still exist the possibility that some parts of the region $\mathcal{L}>\mathcal{L}_{\rm i}$ lie
outside the ellipse and so we still require an enlargement factor but one that is generally smaller than that required by Mukherjee's
method. In Fig \ref{enlargement} evidence estimates are shown calculated with a range of enlargement factors $1.0 < en <
1.1$ as determined for a relatively complex posterior of multiple Gaussians where the analytic $\ln \mathcal{Z} = 5.271$.
The algorithm succesfully converges to the correct value with an enlargement 
factor of $\sim 1.06$.

%%% Comment: Not sure about the close to Gaussian thing, and comment about
%%% non-Gaussian posteriors. Consider the contradiction of a quadratic
%%% posterior. 
% However this ellipse will only provide a good approximation for the
% iso-likelihood curve $\mathcal{L}=\mathcal{L}_{\rm i}$ if the
% posterior is close to Gaussian. For very non-Gaussian posteriors one
% would still require an enlargement factor $>1$.

Computationally it is convenient to draw the random samples not from
this ellipsoidal bound directly but from a unit sphere centred on the
origin. Thus, we construct a transformation matrix $\mathbfss{T} = k
\mathbfss{X}^{\mathbfss{T}}\mathbfss{D}'$ where $\mathbfss{C}' =
\mathbfss{D}'\mathbfss{D}'$ which maps points from the unit sphere into the
ellipse (centred at the origin) so a uniform deviate $\mathbf{x}$ drawn from the sphere
forms a point within the ellipse $\mathbf{y}$ via
\begin{equation}
\mathbf{y} = \mathbfss{T}\mathbf{x} + \mathbf{c}.
\end{equation}

Generating deviates from a uniform $n$-sphere can be most easily achieved by drawing a set of $n$
 Gaussian random numbers $\mathbf{z}$ (via the Box-Muller algorithm), scaling to give a point uniformly distributed
 on the surface of the sphere $\mathbf{z}'= \frac{1}{r}\mathbf{z}$ where $r= \sum z_i^2$ and finally scaling again by a
 deviate drawn from a quadratic distribution between 0 and 1 to generate a point uniformly within the
 sphere. It is worth pointing out that the number of active points $N$ must always be greater than the
 dimensionality of the parameter space in order for the ellipse to be defined unambiguously.

\subsection{Recursive Clustering}

\begin{figure}
 \begin{center}
    	\subfigure[]{
          \includegraphics[width=0.4\columnwidth]{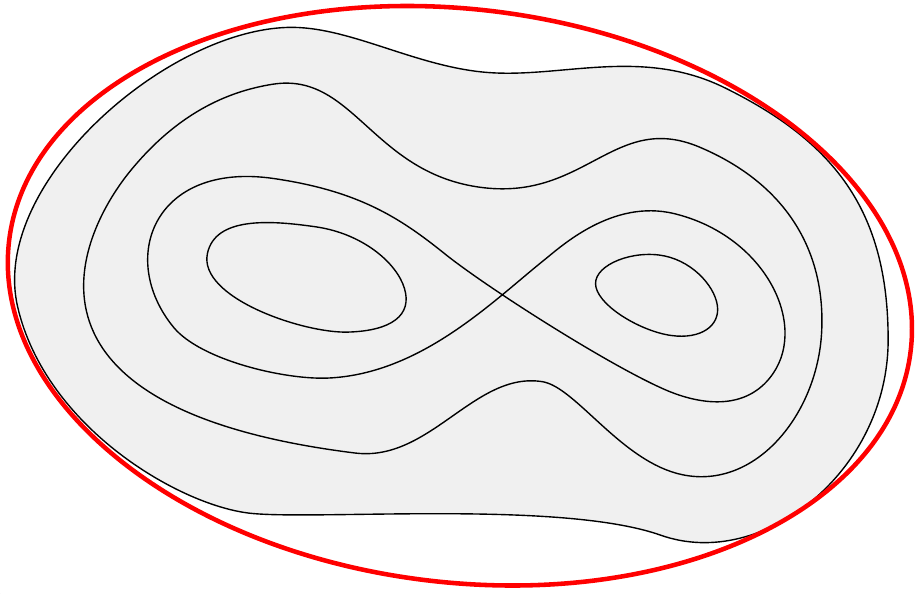}}
	\subfigure[]{
          \includegraphics[width=0.4\columnwidth]{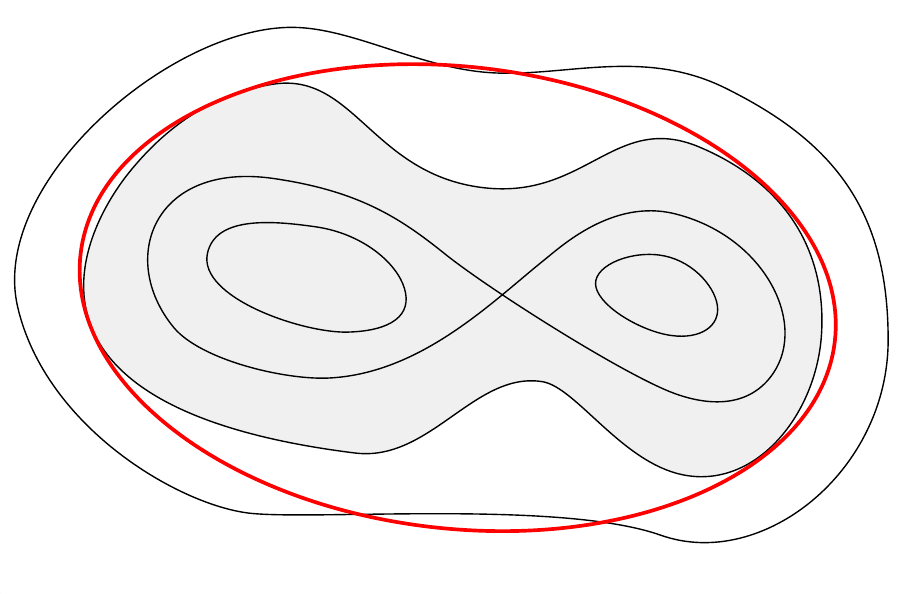}}\\
	\subfigure[]{
          \includegraphics[width=0.4\columnwidth]{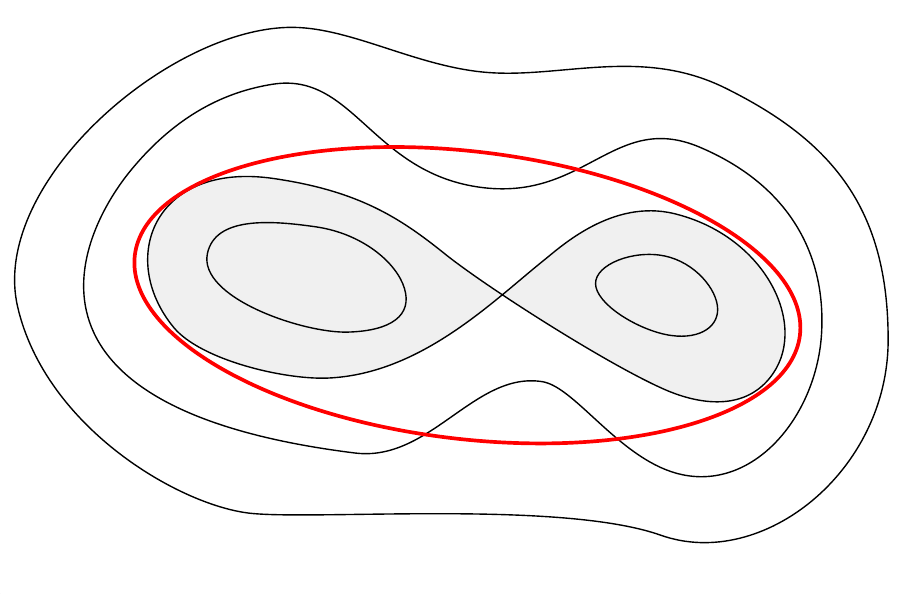}}
	\subfigure[]{
          \includegraphics[width=0.4\columnwidth]{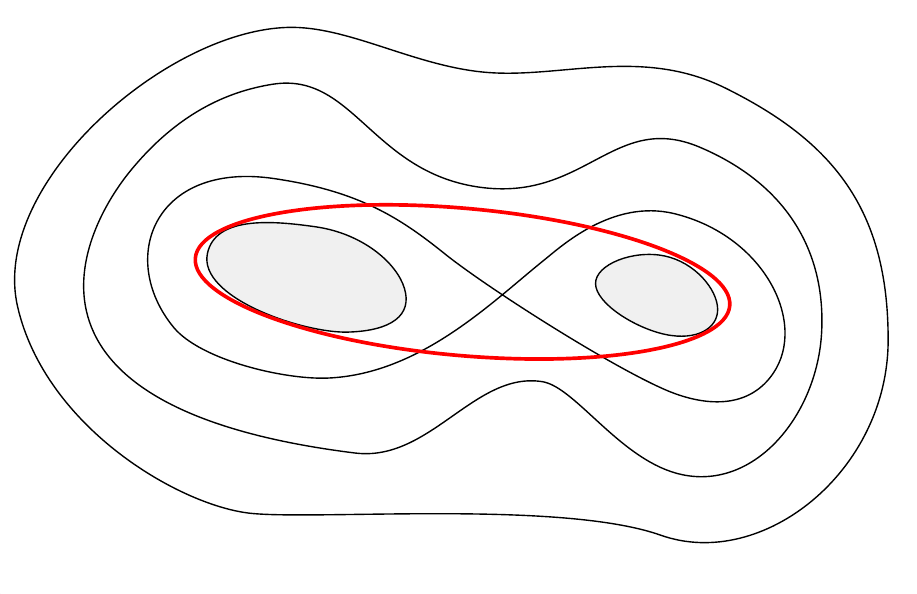}}\\
	 \subfigure[]{
          \includegraphics[width=0.4\columnwidth]{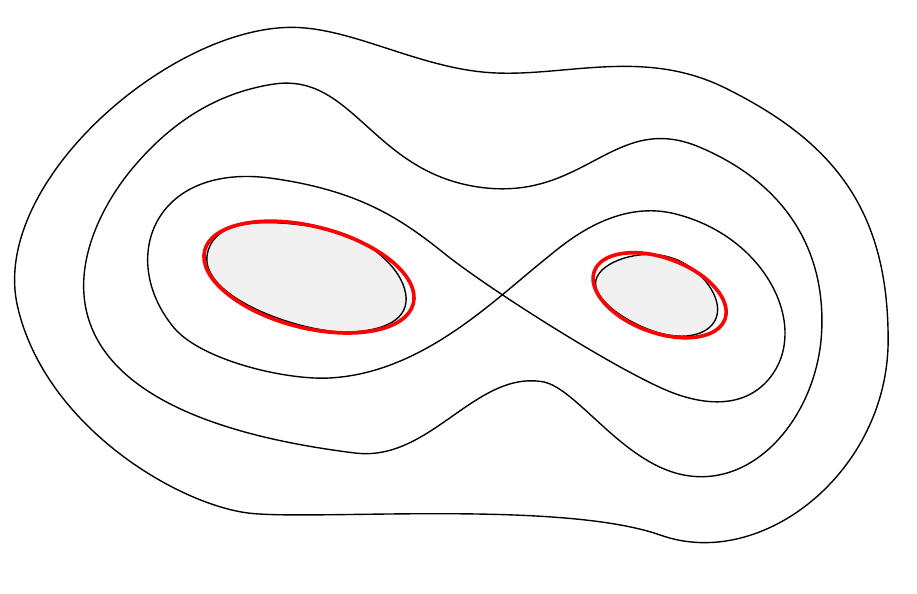}}
        \caption{Cartoon of a simple bimodal case. In (a) we see that
          the ellipsoid represents a good bound to the active
          region. In (b)-(d), as we nest inward we can see that the
          acceptance rate will rapidly decrease as the bound steadily
          worsens. Figure (e) illustrates the solution, we branch and
          sample each clustered region in turn.}
\label{figure2}
\end{center}
\end{figure}

\begin{figure}
 \begin{center}
          \includegraphics[width=0.8\columnwidth]{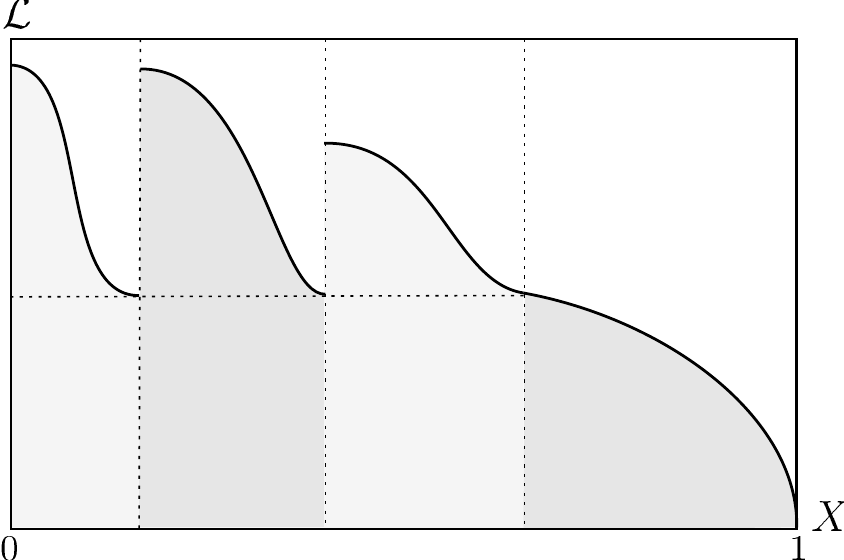}
\caption{View of clustering in $\mathcal{L}-X$ space.}
\label{figure3}
\end{center}
\end{figure}

The usefulness of ellipsoidal sampling is reduced for posteriors that
are not uni-modal. Our extension allows the formation of multiple
ellipsoids centred on individual isolated peaks in the posterior
which can greatly reduce the region of prior sampled and thus
increases sampling efficiency.  The idea is depicted in Fig. \ref{figure2} in a toy
posterior and equivalently in Fig. \ref{figure3} in $X-\mathcal{L}$
space, where upon identifying distinct \emph{clusters} they are
separated and fit by individual ellipsoids. While this does violate
the requirement that $\mathcal{L}(X)$ be a uniquely valued function we
are rescued by the linear nature of the evidence.  It is valid to
consider each cluster individually and sum the contributions provided
we correctly assign the prior masses to each distinct region. Since
our collection of $N$ active points is distributed evenly across the
prior we can safely assume that the number of points within each
clustered region is proportional to the prior mass contained therein.

Identifying clusters in the posterior, in the absence of any
analytical form for the function, is performed by recursive
application of the \emph{K-means} algorithm with $K=2$. This process is
designed to place $N$ data points in an arbitrarily large number of dimensions into 
$K$ clusters. An iterative process the algorithm starts by dividing the set of samples
into two clusters, the mean position of each cluster is then determined and each
subsequent point is assigned to one or other of the two mean positions. The next
iteration updates the mean positions and so on. In this way a set of sampled data can
be searched by the alorithm for clustering centres. 
For a more detailed discussion of the $K$-means process see
for example
see \citealt{Mackay}. 
While our implementation does not require the
user to know beforehand any details about the shape of the posterior,
the algorithm does contain a number of user definable parameters which
can guide the clustering. These parameters define two conditions that must be met 
once two separate clusters have been
identified via K-means. The first is a volume reduction factor
set so that the total volume of the combined clustered
ellipsoids is less than some fraction of the pre-clustered ellipsoid.
The second parameter requires that the clusters are sufficiently separated 
by some distance to avoid
overlapping regions. We have found reasonable values for the former to be $\sim 0.3$
while the latter seperation criteria requires a new ellipse to be at least $1.1$ times
the existing ellipse major axis in each dimension.
If these conditions are met clustering will occur
and the algorithm will search independently within each cluster for
further sub-clusters. This process continues recursively until some
stopping criterion (discussed in Sec.  \ref{implementation:stopping}.)
is met. So long as condition (i) is met we will be guaranteed a
performance improvement as less of the prior needs to be sampled at
each cycle.

%%% This explanation isn't very clear
% Since we used an assessment of the error in assigning prior mass values
% to estimate the error on $\mathcal{Z}$ earlier we have now introduced
% a further source of error by dividing a total $X_i$ among many
% clusters. If the fraction of active points in a single cluster was
% divided among two sub-clusters as: $f_1 = \frac{n_1}{N}$ and $f_2 =
% \frac{n_2}{N}$ where $N=n_1 + n_2$ then the probaility distribution of
% the prior mass fraction $f_1$ is simply binomial:

The division of prior mass between two clusters introduces a further
source of error, and we can no longer simply assess the uncertainty in
$\mathcal{Z}$. To resolve this we must find the probability of the
mass fraction in each cluster. If we have a fraction $f_1$ in cluster
1, with a total number of points $N$, the probability of $n_1$ points
in cluster 1 is simply binomial:
\begin{equation}
P(n_1|f_1, N) = \frac{N!}{n_1!(N-n_1)}f_1^{n_1}(1-f_1)^{N-n_1}
\end{equation}
We can invert this using Bayes' theorem with a uniform prior to yield
$P(f_1|n_1, N)$ and draw samples via MCMC which can then be used, as
in Sec. \ref{section:nested} to generate a variance for each cluster
which can be summed to give the total error.

\subsection{Stopping Criterion}
\label{implementation:stopping}

We wish to stop the calculation on determining the evidence to some
specified accuracy. One way would be to proceed until the evidence
estimated at each replacement changes by less than a specified
tolerance. However, this could for example underestimate the evidence
in cases where the posterior contains any narrow peaks close to its
maximum. \citet{Skilling} provides an adequate and robust condition by
determining an upper limit on the evidence that could be determined
from the remaining set of current active points.

By selecting the maximum likelihood $\mathcal{L}_{\text{max}}$ in the set
of active points one can then safely assume that the largest evidence
contribution that can be made by the remaining portion of the
posterior is $\Delta{\mathcal{Z}}_{\rm i} = \mathcal{L}_{\rm max}X_{\rm i}$ i.e. the
product of the remaining prior mass and maximum likelihood. We choose
to stop when this quantity would no longer change the final evidence
estimate by some user defined factor.

\section{Applications}

To demonstrate the capabilities of clustered nested sampling we now take
three specific examples; the first two involve posteriors of known
functional form so that an analytic evidence can be compared with that
found through our nested sampling algorithm and the final example
demonstrates a cosmological application involving a bi-modal
posterior.

\subsection{Toy model I}
\label{implementation:ExampleI}

\begin{figure}
\begin{center}
%\psfrag{labelz}{$\ln{\mathcal{L}}$}
%\psfrag{labelx}{$\Theta_1$}
%\psfrag{labely}{$\Theta_2$}
\includegraphics[width=\linewidth]{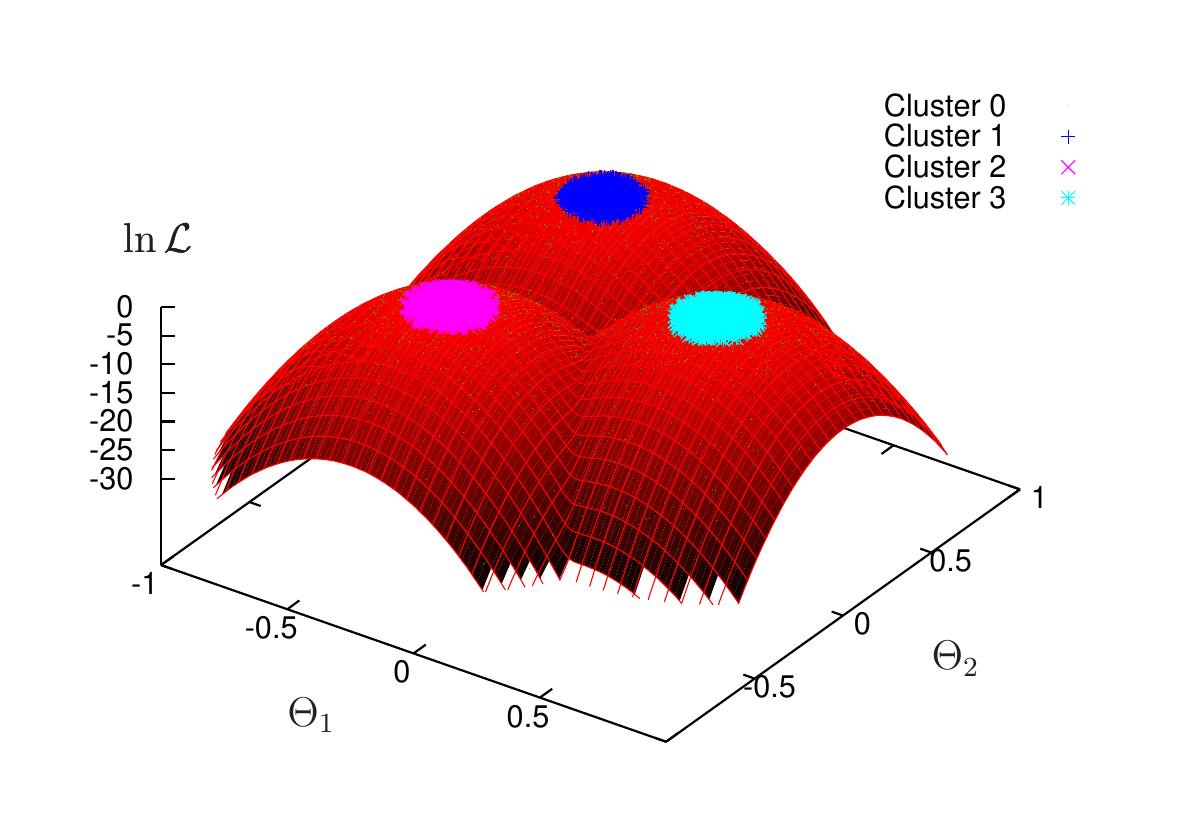}
\caption{Toy Model I: Posterior consisting of 3 Gaussian peaks placed at
  the vertices of an equilateral triangle in the x-y plane at a
  distance of 0.5 units from the origin with $\sigma = 0.1$. The
  clusters found are depicted in colour at each peak.}
\label{ExampleI:3gaussian}
\end{center}
\end{figure}

Our first example posterior consists of 3-Gaussian peaks in a two
dimensional parameter space. The clustering (see
Fig. \ref{ExampleI:3gaussian}) algorithm first divides the space
into cluster 1 and cluster 2+3, runs to the required accuracy on
cluster 1 then returns to the combination of clusters 2+3,
immediately clusters to separate 2+3 then running both successively
to completion.  Table \ref{ExampleI:3gaussian_table} shows that with
and without clustering we can determine the evidence to within 2 \% of
the analytic value. To obtain similar accuracy, however the clustered example
required only 5 \% of the number of likelihood evaluations. Our
calculated errors agree to within 10 \% of the sampling errors based
on 50 repetitions.

\begin{table}
\begin{center}
  \caption{Summary of sampling statistics with and without clustering
    for the 3-Gaussian posterior with a log-evidence, determined
    analytically of 2.813.}
\begin{tabular}{|c||c||c|}
    \hline
Toy Model I: &  Clustering & No Clustering \\
    \hline
 $\ln{\mathcal{Z}}$ & 2.798 & 2.830 \\
 $N_{\rm like}$ & 12,115 & 258,361 \\
 Sampling Error &  0.063 & 0.060\\
 Calculated Error & 0.061 & 0.059\\
    \hline
\end{tabular}
\label{ExampleI:3gaussian_table}
\end{center}
\end{table}

\subsection{Toy model II}
\label{implementation:ExampleII}

\begin{figure}
\begin{center}
%\psfrag{labelz}{$\ln{\mathcal{L}}$}
%\psfrag{labelx}{$\Theta_1$}
%\psfrag{labely}{$\Theta_2$}
\includegraphics[width=\linewidth]{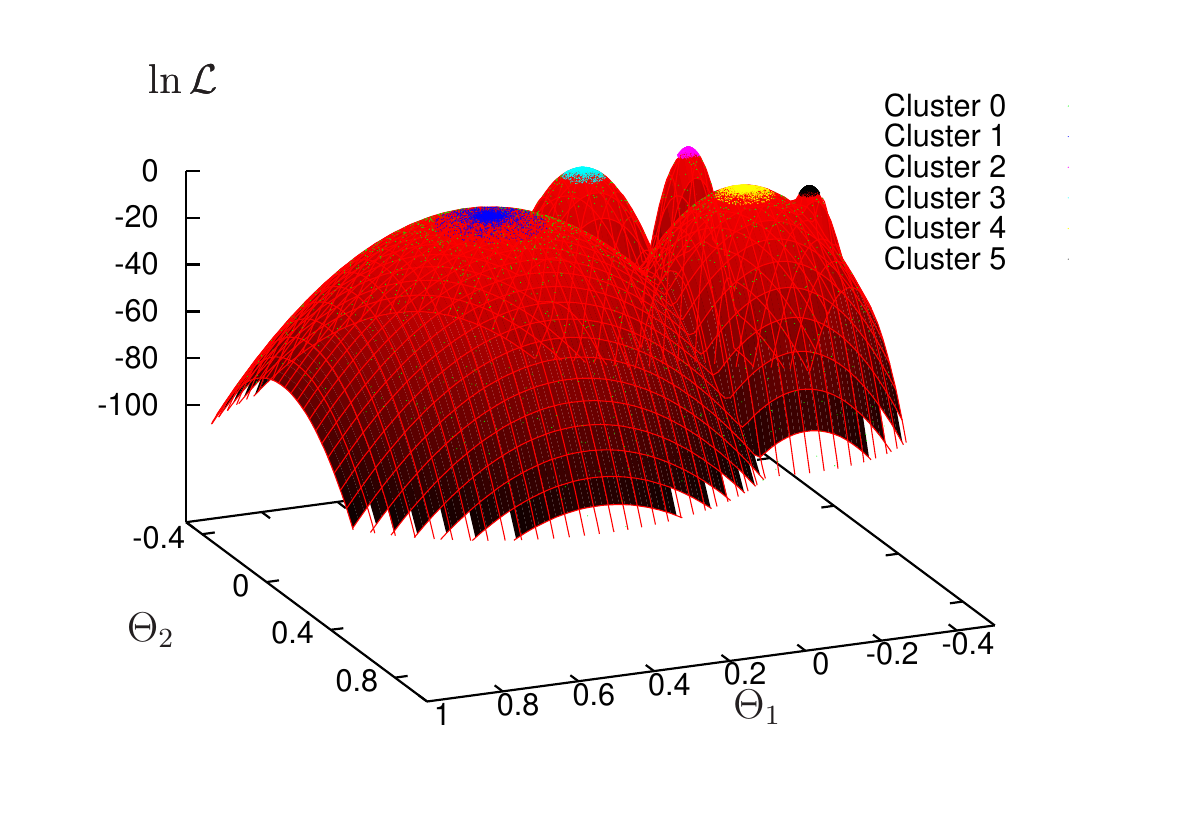}
\caption{Toy Model II: Posterior consisting of 5 Gaussian peaks spaced
  randomly in the x-y plane of varying width and height. The clusters
  found are depicted in colour at each peak.}
\label{ExampleII:comp}
\end{center}
\end{figure}

Posteriors that occur in typical applications may be far more
complicated than that used in Sec. \ref{implementation:ExampleI}. Our
second example uses a combination of five Gaussians of varying width
and height in the x-y plane (see
Fig. \ref{ExampleII:comp}). Clustering occurs naturally in the
expected fashion, producing a improvement in efficiency of at least a
factor of 3 (see Table \ref{ExampleII:comp_table}). Note in particular
the algorithm correctly isolates the slim Gaussian in cluster 5
superimposed on the larger Gaussian at cluster 4.

\begin{table}
\begin{center}
  \caption{Summary of sampling statistics with and without clustering
    for the multi-Gaussian posterior with a log-evidence determined analytically of 5.271.}
\begin{tabular}{|c||c||c|}
    \hline
Toy Model II: &  Clustering & No Clustering \\
    \hline
 $\ln{\mathcal{Z}}$ & 5.296 & 5.230 \\
 $N_{\rm like}$ & 1,016,994 & 3,615,300 \\
 Sampling Error &  0.083 & 0.085\\
 Calculated Error & 0.084 & 0.084\\
    \hline
\end{tabular}
\label{ExampleII:comp_table}
\end{center}
\end{table}

\subsection{Cosmological application}
\label{implementation:ExampleIII}
Recent measurements of the Cosmic Microwave Background (CMB) by WMAP
 indicate a deficiency of power on large scales when
compared with a flat $\Lambda{\rm CDM}$ cosmological model. One possibility
is that this decrease is caused by a feature in the primordial
spectrum of density perturbations, specifically a sharp cutoff on
scales $k \leq 0.0003$ Mpc$^{-1}$. Many analyses to date have examined
various forms for this cutoff (for example \citealt{Efstathiou};
\citealt{Bridle}; \citealt{Souradeep}; \citealt{Lasenby}), here we
will examine a model with favourable likelihood as determined by
\citet{Sinha} using WMAP 1-year TT data \citep{WMAP1}. The posterior
derived for this model/data combination contains substantial
multi-modality in the parameters defining the spectrum.
 
\begin{figure}
\begin{center}
%\psfrag{zlabel}{$\ln{\mathcal{L}}$}
%\psfrag{xlabel}{$k_c$}
%\psfrag{ylabel}{$\alpha$}
\includegraphics[width=\linewidth]{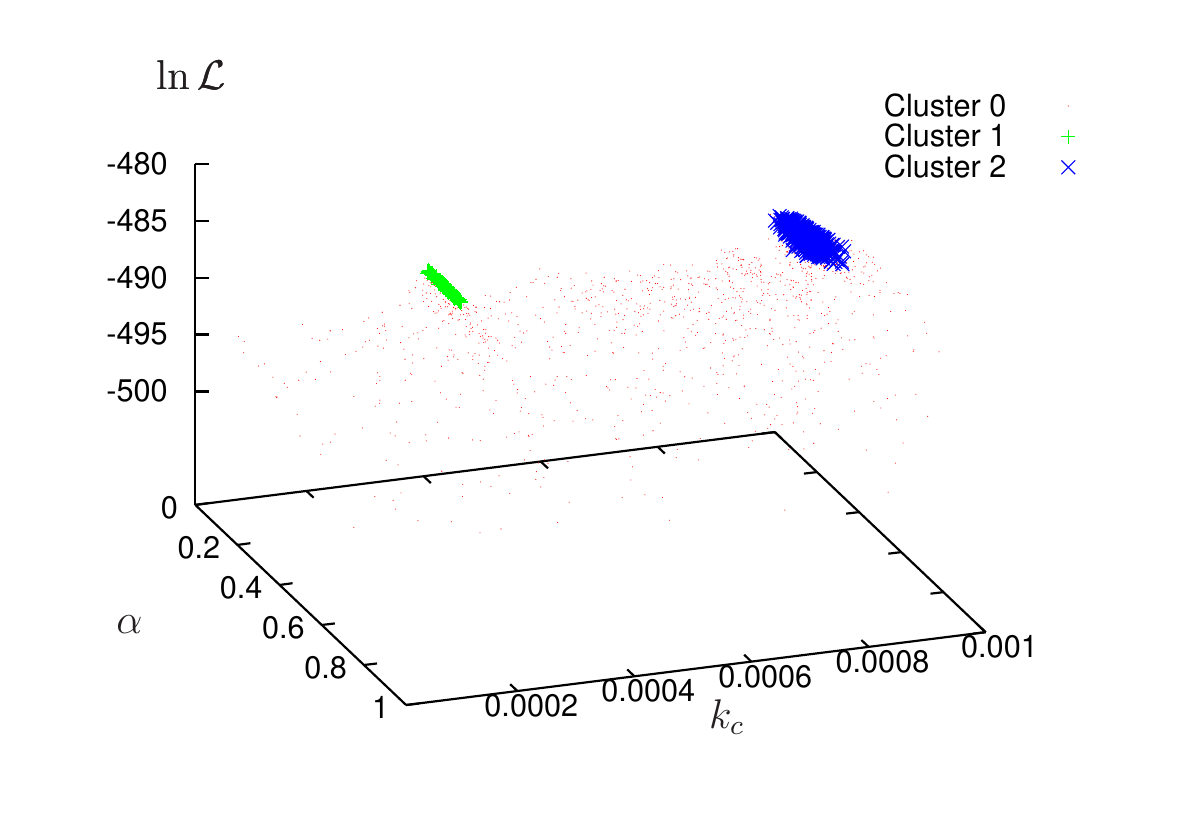}
\caption{Cosmological Application: Set of replacement samples in the cutoff scale $k_c$ and
  $\alpha$ with clustered regions shown.}
\label{ExampleIII:bimodal}
\end{center}
\end{figure}

\begin{figure}
\begin{center}
%\psfrag{zlabel}{$\mathcal{L}$}
%\psfrag{xlabel}{$k_c$}
%\psfrag{R}{$R_{\ast}$}
\includegraphics[width=\linewidth]{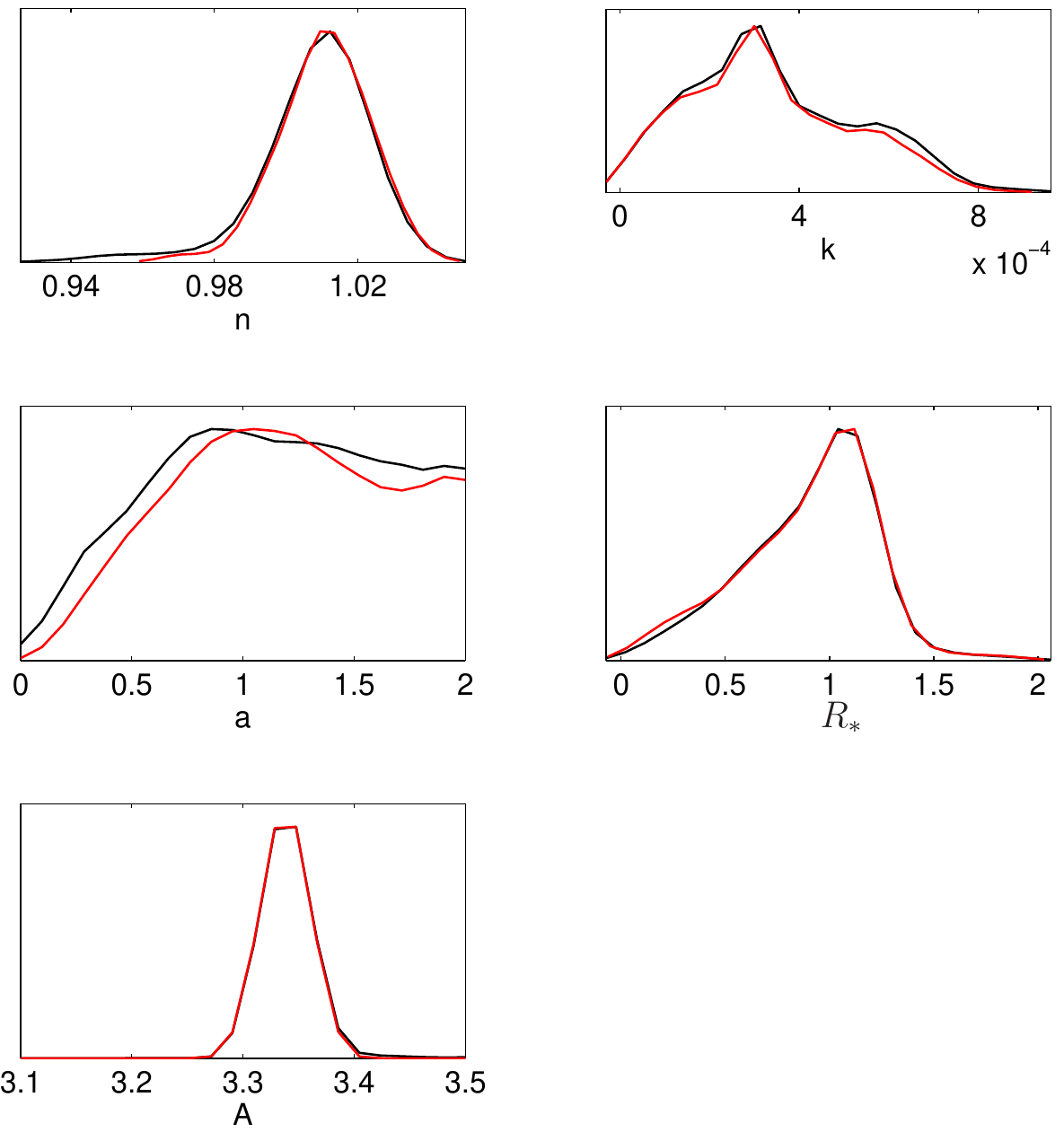}
\caption{Cosmological Application: 1-D marginalised posteriors for model spectral
  parameters, comparing those obtained via MCMC (red) and nested
  sampling (black).}
\label{ExampleIII:posterior}
\end{center}
\end{figure}

Physically a sharp cutoff can be generated by a kink in the
inflationary potential \citep{Starobinsky}, followed by a `bump' and
damped oscillations at larger $k$. This effect can be modelled via a
transfer function applied to any underlying spectrum $P(k) = P_0(k)
T^2(y, R_{\ast})$ where $y=k/k_{\ast}$ and
\begin{multline}
T^2(y, R_{\ast})=1-3(R_{\ast}-1)\frac{1}{y}\left[ \left(1-\frac{1}{y^2}\right)\sin 2y +
\frac{2}{y}\cos 2y\right] \\
+ \frac{9}{2}(R_{\ast})^2\frac{1}{y^2}\left(1+\frac{1}{y^2}\right) \\
\left[ 1+\frac{1}{y^2}+\left(1-\frac{1}{y^2}\right)\cos 2y - \frac{2}{y}\sin 2y \right].
\end{multline}
$R_{\ast}$ is the ratio of inflationary potential and scalar field
where $k_{\ast}$ is the cutoff wave-vector. The best fit model as found
by \citet{Sinha} was a slight variant on this with an exponential
cutoff such that the underlying spectrum is a single index spectrum
with exponential decay, governed by a factor $\alpha$:
\begin{equation}
P(k) = A(1-e^{(0.75y)^{\alpha}}) k^{n-1} T^2.
\end{equation} 
We have restricted the analysis to the set of parameters defining the
spectrum only ($n_s$, $k_c$, $\alpha$, $R_{\ast}$, $A_s$) and fixed the
remaining parameters to their best fit values as determined via MCMC using the {\sc CosmoMC} package
\citep{cosmomc} (namely $\Omega_b h^2 = 0.0243$, $\Omega_c h^2$ = 0.115,
$\theta = 1.04$ and $\tau = 0.22$). The results of \citet{Starobinsky} suggest that the posterior
is bimodal in the subspace $(\alpha, k_c)$, andso this problem provides an ideal example of the applicability of clustering.
The algorithm correctly identifies both peaks and clusters accordingly
(see Fig. \ref{ExampleIII:bimodal}). The marginalised posteriors for each parameter are shown in Fig. \ref{ExampleIII:posterior}

The maximum likelihood point in
the marginalised posterior of $k_c$ is clearly centred at $0.0003$
Mpc$^{-1}$ (see Fig. \ref{ExampleIII:posterior}) with a secondary peak
also visible at approximately $0.0006$ Mpc$^{-1}$. These features
provide an ideal example of the applicability of clustering. The
algorithm correctly identifies both peaks and clusters accordingly
(see Fig. \ref{ExampleIII:bimodal}).

We completed 10 separate evaluations of the evidence via the three
most common methods for comparison. Table
\ref{ExampleIII:bimodal_table} summarises the evidences obtained,
their associated errors in sampling (i.e. the standard error over the
10 runs), the calculated error estimated via the method described in
Sec. \ref{implementation:stopping} and the number of likelihood calls
made per individual run. Both nested sampling methods provide
significant improvements in $N_{\rm like}$ over thermodynamic
integration \footnote{Note that although thermodynamic integration
  took only twice the number of likelihood calls its associated
  sampling error is still significantly larger.}. With clustering
however we see a further performance increase by more than a factor of
2 for similar statistical uncertainties. Moreover, the sampling error
and our calculated error show extremely good agreement, eliminating
the need to perform more than one evidence calculation and thus
further reducing the computational cost by an order of magnitude.

\begin{table}
\begin{center}
  \caption{Sampling statistics for the Starobinsky model via nested
    sampling (both clustered and without) and thermodynamic
    integration (TI).}
\begin{tabular}{|c||c||c||c|}
    \hline
Example III: &  Clustering & No Clustering & TI\\
    \hline
 $\ln{\mathcal{Z}}$ & 494.65 & 494.45 & 493.76\\
 $N_{\rm like}$ & 96,720 & 250,523 & $>$ 400,000\\
 Sampling Error &  0.29 & 0.31 & 0.87 \\
 Calculated Error & 0.30 & 0.29 & --\\
    \hline
\end{tabular}
\label{ExampleIII:bimodal_table}
\end{center}
\end{table}

\section{Conclusions}
We have demonstrated the applicability of our clustered nested sampling algorithm  to multi-modal posterior distributions in
both simulated `toy' models and a cosmological example based on the Starobinsky primordial power spectrum --in which a
substantial performance increase was noted. Although we have examined here a multi-modal cosmological
example, our sampler is by no means restricted to such use. In fact we feel an important feature is the ability to
determine the variance of the final evidence value, for any model, without calculating the sample variance over repetitive runs. 
This alone reduces the computational load by at least an order when compared with previous methods.

\section*{Acknowledgements}

This work was carried out largely on the COSMOS UK National Cosmology
Supercomputer at DAMTP, Cambridge and we would like to thank S. Rankin
and V. Treviso for their computational assistance. The authors would like to thank Andrew Liddle, David MacKay and Faran Ferhoz for useful
discussions. JRS was supported by STFC. MB was supported
by a Benefactors Scholarship at St. John's College, Cambridge and an
Isaac Newton Studentship.

\appendix

\label{lastpage}

\end{document}